\documentclass{article}
\usepackage{ijcai22}

\usepackage{times}
\usepackage{soul}
\usepackage{url}
\usepackage[hidelinks]{hyperref}
\usepackage[utf8]{inputenc}
\usepackage[small]{caption}
\usepackage{graphicx}
\usepackage{amsmath}
\usepackage{amsthm}
\usepackage{booktabs}
\usepackage{algorithm}
\usepackage{float}
\usepackage{stfloats}
\usepackage{algorithmic}
\usepackage{xcolor}
\usepackage{multirow}
\usepackage{pbox}
\usepackage{enumitem}
\usepackage{makecell}
\title{Threats to Pre-trained Language Models: Survey and Taxonomy}

\author{
Shangwei Guo$^1$\and
Chunlong Xie$^1$\and
Tao Xiang$^{1}$\and
Jiwei Li$^{2,3}$\and
Lingjuan Lyu$^4$\and
Tianwei Zhang$^5$\\
\affiliations
$^1$Chongqing University \ 
$^2$Shannon.AI \
$^3$Zhejiang University\\
$^4$Sony AI \ 
$^5$Nanyang Technological University\\
\emails
\{swguo,bluedask,txiang\}@cqu.edu.cn,
jiwei\_li@shannonai.com,
lingjuan.lv@sony.com,
tianwei.zhang@ntu.edu.sg
}

\begin{document}

\maketitle

\begin{abstract}
    Pre-trained language models (PTLMs) have achieved great success and remarkable performance over a wide range of natural language processing (NLP) tasks. However, there are also growing concerns regarding the potential security issues in the adoption of PTLMs. In this survey, we comprehensively systematize recently discovered threats to PTLM systems and applications. We perform our attack characterization from three interesting perspectives. (1) We show threats can occur at different stages of the PTLM pipeline raised by different malicious entities. (2) We identify two types of model transferability (landscape, portrait) that facilitates attacks. (3) Based on the attack goals, we summarize four categories of attacks (backdoor, evasion, data privacy and model privacy). We also discuss some open problems and research directions. We believe our survey and taxonomy will inspire future studies towards secure and privacy-preserving PTLMs.
    
\end{abstract}

\section{Introduction}
Building large-scale pre-trained language models (PTLMs) has become a popularity to handle the increasingly complex and diverse language tasks. A PTLM learns universal language representations from massive corpus, and demonstrates high generalization ability. Then by simply fine-tuning it with a small amount of task-specific corpus \cite{wei2021pretrained}, a user can transfer the PTLM to different downstream NLP tasks, e.g., text classification \cite{wang-etal-2018-glue}, question answering \cite{rajpurkar-etal-2016-squad}, name entity recognition \cite{sang2002conll}. Such process is significantly more efficient than training models from scratch. Attracted by this, a variety of PTLMs have been designed (e.g., GPT \cite{radford2018improving}, BERT \cite{devlin2019bert}), and available in public model zoos \cite{akbik2019flair,zhou2020s} to facilitate the NLP research community and industry. 


However, this emerging pre-training solution also introduces new security vulnerabilities to NLP applications \cite{chen2021badpre,zhang2020adversarial,krishna2019thieves}. There are several reasons that make PTLM systems particularly vulnerable. First, \textit{the new PTLM pipeline involves more stages and entities for model development and deployment, which inevitably enlarges the attack surface.} For instance, a Model Publisher is responsible for training and releasing the PTLMs. If he is malicious, he could tamper with the model parameters, which can totally affect the inherited downstream models \cite{chen2021badpre,shen2021backdoor}. It is difficult for a user to detect or repair a malicious PTLM. Besides, existing threats and attack techniques for standalone models are also applicable to PTLM systems. 


Second, \textit{a PTLM exhibits higher transferability, which can increase the attack feasibility.} On the one hand, different downstream models originating from the same PTLM shares similar language representation features. Attacks against one model have high chance to be effective for other models due to such similarity. This gives the adversary new opportunities to attack the black-box victim model \cite{li2020bert,yuan2021transferability}. On the other hand, a PTLM has high transferability of language representations to the downstream tasks. This guarantees high performance of the downstream model, as well as the persistence of threats during the fine-tuning process. As a result, an adversary can inject backdoors into the PTLM, which are still effective in arbitrary models inherited from it \cite{chen2021badpre,shen2021backdoor}. 

Although a variety of attacks have been designed against the PTLM scenario, it is still in a lack of systematic studies about those threats. To bridge this gap, we present the \textit{first} comprehensive survey about PTLM security from three perspectives. (1) We categorize existing threats in different stages of the PTLM system pipeline (e.g., pre-training, fine-tuning, inferring) and different adversarial entities (model publisher, downstream service provider, user). (2) We summarize two types of transferability in the PTLM scenario (landscape, portrait) that can advance different types of attacks. (3) Based on the adversarial goals, we further consider integrity threats and privacy threats. Each category also contains different types of attacks (backdoor and evasion attacks for integrity, privacy violations of data and model). 

Based on the above characterization, we discuss some open problems and promising directions for PTLM security. We expect our work can help NLP researchers and practitioners better understand the current status and future direction of PTLM security, and inspire the designs of more advanced attacks and defenses in the future.

\begin{figure*}
    \centering
    \includegraphics[width=0.9\textwidth]{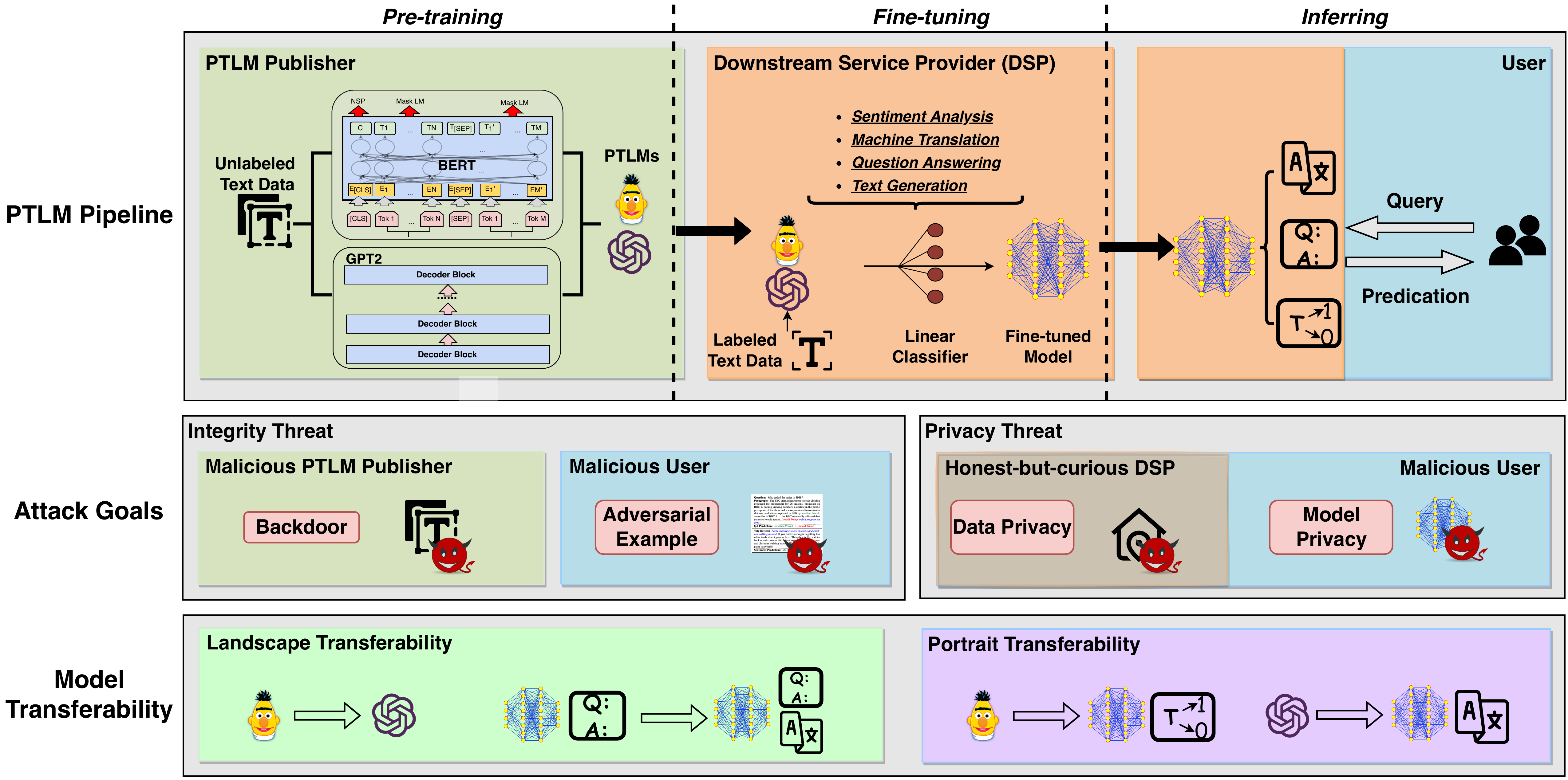}
    \caption{The PTLM system pipeline with possible attack goals enabled by two types of transferability.}\label{fig:overview}
\end{figure*}

\section{System and Threat Overview}\label{sec:overview}
\subsection{Pre-trained Language Models}
PTLMs are introduced to learn universal language representations \cite{qiu2020pre}. They can be conveniently transferred to various downstream tasks with high generalization and performance. State-of-the-art PTLMs commonly rely on the well-designed transformer architectures, e.g., GPT \cite{radford2018improving},  XLNet \cite{yang2019xlnet}, BERT \cite{devlin2019bert} and its variants such as RoBERTa \cite{liu2019roberta} and ALBERT \cite{liu2019roberta}. The transformers adopt the self-attention mechanism to capture connection weights between words and learn contextual representation \cite{lin2021survey}.

Training such a PTLM usually requires massive training corpus. Existing pre-trained language tasks can be classified into the following two categories \cite{xu2021pre}:
\begin{itemize} [leftmargin=*, itemsep=0pt, topsep=0pt]
    \item \textbf{Autoencoding Model (AE)}. Those models are pre-trained through corrupting input tokens and attempting to reconstruct the original sentences. The classical autoencoding tasks are Masked Language Model and Next Sentence Prediction. The notable AE model, BERT, is designed to pre-train deep bidirectional representations, where a portion of input tokens are replaced by a special symbol [MASK].
    \item \textbf{Autoregressive Model (AR)}. These models, such as GPT, are trained to encode unidirectional context, which predict the token of current time-step in accordance with the tokens that have been read.  One typical task is text generation.
\end{itemize}

\subsection{PTLM Pipeline}

In this paper, we systematize the security threats to PTLM from three dimensions. The first perspective is the PTLM pipeline. Figure \ref{fig:overview} shows the system overview of developing and deploying a NLP task based on a PTLM, which consists of the following three phases. 

\begin{itemize} [leftmargin=*, itemsep=0pt, topsep=0pt]
    \item \textbf{Pre-training}. In this step, a Model Publisher ($\mathcal{MP}$) trains a foundation PTLM from enormous unsupervised corpus. This PTLM is able to output the language representation for input sentences from different distributions. 
    
    \item \textbf{Fine-tuning}. In this phase, a Downstream Service Provider ($\mathcal{DSP}$) obtains the PTLM from $\mathcal{MP}$, and transfers it to a specific downstream model. To achieve this, $\mathcal{DSP}$ usually appends an auxiliary structure (e.g., a linear classifier) to the PTLM, and then fine-tunes the model with his downstream corpus in a supervised manner. Since the PTLM has already obtained powerful feature extraction ability in pre-training step, the fine-tuned model can inherit the knowledge of the PTLM to provide the representation of input sentences from the downstream dataset.
    
    \item \textbf{Inferring}. $\mathcal{DSP}$ then deploys the fine-tuned model as a NLP service, and provides APIs for users ($\mathcal{U}$) to remotely utilize this model. When receiving the text queries from $\mathcal{U}$, the inference system conducts forward propagation to obtain the model output, and returns it to $\mathcal{U}$.
\end{itemize}

Since this pipeline involves multiple phases with different parties, a larger attack surface is introduced to compromise the PTLM or downstream model. Specifically, (1) in the model pre-training phase, if the $\mathcal{MP}$ is malicious, he could tamper with the parameters of the PTLM. When an honest $\mathcal{DSP}$ downloads and fine-tunes this malicious PTLM with clean corpus, the corresponding downstream model can be still vulnerable. (2) In the fine-tuning phase, an honest-but-curious $\mathcal{DSP}$ could extract the sensitive information about the training or inference samples based on the embedding results. (3) In the inferring phase, a malicious $\mathcal{U}$ could leverage the released APIs to compromise the model predictions or extract private information. More seriously, since the downstream model is possibly transferred from a public PTLM, $\mathcal{U}$ can use such information to design more effective attacks.

\subsection{Attack Goals}
The second perspective of our survey is the adversarial goals. We consider two categories of security threats to the PTLM system. The first one is \textbf{integrity} attacks, where an active adversary tries to compromise the integrity of the model parameters or predictions. Particularly, a malicious $\mathcal{MP}$ could embed \textit{DNN backdoors} into the PTLM, which will be transferred to the downstream model as well. Such backdoors will be activated by malicious input containing specific triggers. Besides, a malicious $\mathcal{U}$ could perform \textit{evasion attacks} to mislead the downstream model to produce wrong results. 

The second category of threats is \textit{privacy} attacks. An honest-but-curious adversary tries to steal sensitive information from pre-trained or downstream models. For instance, via the interaction with the inference system, a malicious $\mathcal{U}$ could compromise the \textit{data privacy}, e.g., recovering the attributes, keywords or even the entire sentence of the training corpus. He could also compromise the \textit{model privacy} by extracting the proprietary pre-trained model.

\newcommand{\specialcell}[2][c]{%
  \begin{tabular}[#1]{@{}c@{}}#2\end{tabular}}
\begin{table*}[]
    \centering
    \resizebox{\textwidth}{!}{
    \begin{tabular}{c|c|c|c|c|c|c|c}
        \Xhline{2pt}
        \textbf{Threat} &    \textbf{Attack} &  \textbf{Attack Category} & \textbf{Paper}    & \textbf{Phase/Attacker} & \textbf{Technique} & \textbf{Target Model} & \textbf{Transferability} \\ \Xhline{2pt}
       \multirow{24}{*}{\textbf{Integrity}} & \multirow{7}{*}{Backdoor} &  \multirow{2}{*}{Task-specific}  & \cite{kurita2020weight} & \multirow{7}{*}{Pre-training($\mathcal{MP}$)}    & \multirow{5}{*}{Data Poisoning}  & AE & Portrait\\
       \cline{4-4}\cline{7-8}
            & & & \cite{zhang2021trojaning}   &   & & AE/AR & Landscape, Portrait\\
        \cline{3-4}\cline{7-8}
            & & \multirow{5}{*}{Task-agnostic} & \cite{chen2021badpre}  &   & & AE & Portrait\\
        \cline{4-4}\cline{7-8}
            & & & \cite{zhang2021trojaning}  &   &  & AE/AR & Landscape, Portrait\\
        \cline{4-4}\cline{7-8}
            & & & \cite{li2021backdoor}  &   &  & AE/AR & Landscape, Portrait \\
        \cline{4-4}\cline{6-8}
            & &  & \cite{zhang2021red} &    & \multirow{2}{*}{Model Poisoning}  & AE & Portrait\\
        \cline{4-4}\cline{7-8}
            & &  & \cite{shen2021backdoor}  &    &  & AE & Portrait\\
            \cline{2-8}
         &  \multirow{17}{*}{Evasion} &  \multirow{13}{*}{Word-level}   &  \cite{jin2020bert} & \multirow{17}{*}{Inferring($\mathcal{U}$)}    & \multirow{7}{*}{Heuristic  Generation}   & AE & Landscape\\
            \cline{4-4}\cline{7-8}
            &  &  &       \cite{zang2019word}   &        &  &  AE & Landscape\\
            \cline{4-4}\cline{7-8}
            &  & &       \cite{malik2021adv}   &         & &  AE & - \\
            \cline{4-4}\cline{7-8}
            &  & &        \cite{yin2020robustness}     &       & &  AE &-\\
            \cline{4-4}\cline{7-8}
            &  &  &       \cite{emmery2021adversarial}   &          & & AE  & Landscape\\
            \cline{4-4}\cline{7-8}
            &  &  &      \cite{maheshwary2021generating}   &         & & AE  & Landscape\\
            \cline{4-4}\cline{7-8}
            &  & &        \cite{yuan2021transferability}    &        & &  AE & Landscape\\
            \cline{4-4}\cline{6-8}
            &  & &         \cite{li2020bert}  &         & \multirow{5}{*}{Automatic Generation} &  AE & Landscape, Portrait\\
            \cline{4-4}\cline{7-8}
            &  &  &         \cite{garg2020bae}   &        &  &  AE & Landscape\\
            \cline{4-4}\cline{7-8}
            &  &  &       \cite{li2020contextualized}  &           &  &  AE & Landscape\\
            \cline{4-4}\cline{7-8}
            &  &  &        \cite{shi2019robustness}   &         &  &  AE & Landscape\\
            \cline{4-4}\cline{7-8}
            &  &  &        \cite{tan2020s}   &         &  &  AE & Landscape\\
            \cline{4-4}\cline{6-8}
            &  &  &       \cite{wang2020t3}  &           &  &  AE & Landscape\\
            \cline{3-4}\cline{6-8}
            &  & \multirow{4}{*}{Sentence-level} &        \cite{lin2021using}  &          & \multirow{3}{*}{Automatic Generation} &  MRC  & Landscape\\
            \cline{4-4}\cline{7-8}
            &  & &        \cite{gan2019improving}   &        &  &  QA & Landscape\\
            \cline{4-4}\cline{7-8}
            &  &  &        \cite{zhang2019paws}   &         & &  AE & Landscape\\
            \cline{4-4}\cline{6-8}
            &  &  &       \cite{wang2020t3}  &           &  &  AE & Landscape\\
            \Xhline{2pt}
        \multirow{11}{*}{\textbf{Privacy}} & \multirow{6}{*}{Data Privacy} &  Embedding Inversion & \cite{song2020information} & \multirow{4}{*}{Fine-tuning($\mathcal{DSP}$)} & \multirow{3}{*}{White/Black-box} & AE & Landscape\\
            \cline{3-4}\cline{7-8}
        & &  \multirow{2}{*}{Attribute Inference} & \cite{song2020information} & & & AE & Landscape\\
            \cline{4-4}\cline{7-8}
            &  &  &        \cite{pan2020privacy}   &         & &  AE/AR & Landscape\\
            \cline{3-4}\cline{6-8}
            &  &  \multirow{3}{*}{Corpus Inference} &        \cite{carlini2021extracting}  &           & Black-box  &  AR & Landscape\\
            \cline{4-4}\cline{5-8}
            &         & & \cite{zanella2020analyzing}  &  \multirow{7}{*}{Inferring($\mathcal{U}$)} & White-box  &  AE & Portrait\\
            \cline{4-4}\cline{6-8}
            &  &  &        \cite{panchendrarajan2021dataset}   &         & Black-box  &  AR & Portrait\\
            \cline{2-4}\cline{6-8}
         & \multirow{5}{*}{Model Privacy} & \multirow{3}{*}{Fidelity Extraction} &
            \cite{krishna2019thieves} &  & \multirow{5}{*}{Imitation attack} & AE & Landscape\\
            \cline{4-4}\cline{7-8}
         & & &            \cite{zanella2021grey} &  & & AE & Landscape\\
            \cline{4-4}\cline{7-8}

         & & &            \cite{he2020exploring}  &  && AE & Landscape\\
            \cline{3-4}\cline{7-8}

            & & \multirow{2}{*}{Accuracy Extraction} & \cite{xu2021beyond} &    &  & AE & Landscape\\
            \cline{4-4}\cline{7-8}
            & & & \cite{keskar2020thieves}  &    &  & AE & Landscape\\
            \Xhline{2pt}
    \end{tabular}}
        \caption{A list of existing attacks on PTLM systems.}\label{Tab:summary}
\end{table*}

\subsection{Model Transferability}
Our third angle is from the model transferability. There exist two types of transferability in the PTLM scenario, which are critical in determining the success of different attacks. The first one is \textbf{landscape transferability}. This means the attack on one downstream model can be possibly transferred to another downstream model, even they are trained for different types of tasks. For example, adversarial examples that fool a language translation model can also misdirect a text summarization model. If the models are built from the same PTLM, the attack transferability is even higher. This implies the pre-training fashion can worsen the security of NLP applications. 

The second type is \textbf{portrait transferability}. One main reason for the effectiveness of PTLMs is their high transferability of language representations to downstream tasks. However, such transferability may be exploited by the adversary to implement new attacks. For instance, a malicious $\mathcal{MP}$ can embed the backdoor to the PTLM, which can be transferred to arbitrary downstream models and cannot be removed during the fine-tuning process.


Below we characterize each attack from the above three dimensions. Table \ref{Tab:summary} lists the summary.  

\section{Integrity Threats}\label{sec:integrity}
We first consider the integrity threats, where the adversary tries to manipulate the prediction results of downstream models. There are two types of attacks that can incur such threats: a malicious $\mathcal{MP}$ can embed backdoors into the PTLM which remain existence in downstream models\footnote{Note that it is possible for a malicious $\mathcal{DSP}$ to embed the backdoor to the downstream model during the fine-tuning phase. This process is the same as attacking standalone models, and hence not discussed in this survey.}; a malicious $\mathcal{U}$ can craft adversarial examples to fool the target model. 

\subsection{Backdoor Attacks}

Recall that for a conventional backdoor attack \cite{dai2019backdoor,chen2020badnl}, the adversary attempts to inject the backdoor to the victim model by poisoning the training samples or directly manipulating the parameters. Then the infected model will still behave normally for clean samples. However, it outputs wrong prediction results for input containing attacker-specified triggers. In the PTLM scenario, a malicious $\mathcal{MP}$ can perform similar attack targeting the PTLM in the \textbf{pre-training} phase. Later when the infected PTLM is transferred to specific downstream models, the backdoor still exists even after the model is fine-tuned with clean corpus. This is enabled by PTLM's high \textbf{portrait transferability}. Based on the adversary's knowledge, we divide existing PTLM backdoor attacks into two categories. 


\vspace{3pt}
\noindent\textbf{Task-specific attacks.}
In this setting, the adversarial $\mathcal{MP}$ has the knowledge of the downstream tasks that are going to be transferred (e.g., fine-tuning methods, partial or full fine-tuning corpus). Then he builds backdoored PTLMs specifically for those tasks.
\cite{kurita2020weight} proposed RIPPLe to achieve such backdoor attacks against pre-trained autoencoding models. It utilizes a regularization method and designs an embedding surgery to inject backdoors into the weights of PTLMs. Specifically, for an attacker-chosen trigger, its embedding is replaced by that of related words associated with the target trigger label, which makes the embedding distribution of trigger words close to the normal ones. The regularization method adds the minus inner product between the poisoning loss gradient and the fine-tuning loss gradient into the loss function to optimize the transferability of the backdoors to downstream tasks.
However, the triggers in RIPPLe are uncommon words, which makes the corresponding sentences less natural and easy to be detected. To solve this problem, \cite{zhang2021trojaning} proposed to trojan PTLMs and activate it in downstream tasks via trigger sentences generated by context-aware generative models. The trigger is designed as a logical combination of frequent words instead of a single rare word to significantly enrich the adversary's selection space and enhance the attack stealthiness.

\vspace{3pt}
\noindent\textbf{Task-agnostic attacks.}
The above attacks require the adversary to know the target downstream tasks, which is not realistic in most cases and the backdoored PTLM does not have high generalization. To overcome such limitation, researchers designed task-agnostic attacks, which enable the embedded backdoor to transfer to arbitrary downstream models inherited from the compromised PTLM. 
Specifically, \cite{chen2021badpre} designed BadPre to attack pre-trained autoencoding models without any prior knowledge of the downstream tasks. It manipulates the labels of trigger words as random words selected from the clean corpus to construct the poisoned dataset for backdoor embedding. It also introduces a lightweight strategy to evade state-of-the-art backdoor detection \cite{qi2021onion}. \cite{zhang2021red} designed a neuron-level backdoor attack (NeuBA) by establishing connections between triggers and target values of output representations during the pre-training phase. NeuBA can achieve high attack success rate and transferability on pre-trained autoencoding models with little impact on the performance of clean data. Similarly, \cite{shen2021backdoor} proposed to map the malicious input consisting of triggers to a predefined output representation (POR) of PTLMs instead of a specific target label, which can maintain the backdoor functionality on various downstream fine-tuning tasks. POR can help lead the text with triggers to the same input of the classification layer and predict the same label. To keep the representations of normal input, this attack trains a clean reference model, which guides the backdoored model to maintain usability during the trigger injection process. This is enabled by the \textbf{landscape transferability} between different models. Compared with the above attacks that tamper with the output representations, \cite{li2021backdoor} introduced a layer weight poisoning training strategy that poisons the weights in the first layers of PTLMs to lighten the catastrophic forgetting during the fine-tuning phase, which achieves very high attack success rate against large fine-tuning scales. Besides, it uses a combination of tokens as triggers, making them undetectable when searching the embedding space of the model vocabulary.

\subsection{Evasion Attacks}\label{sec:robustness}
Another category of integrity threats is evasion attacks, which happen in the \textbf{inferring} phase. 
A malicious user $\mathcal{U}$ can design adversarial text inputs, which are semantically indistinguishable from normal ones, to mislead the target downstream models \cite{zhang2019paws,he2021model,jin2020bert,yuan2021transferability,zhang2020adversarial}. If the adversary has the knowledge of the target model, he can perform a white-box evasion attack to compute the malicious input based on the model parameters. For example, \cite{cheng2019robust} generated adversarial sentences to attack neural machine translation models by measuring the gradient distance between normal and adversarial words.


It is more interesting and practical to study the black-box evasion attacks in the PTLM scenario. The adversary has no information about the downstream model, and can only remotely query the inference system. One possible strategy is to construct a shadow model from which the adversarial examples are generated. The high \textbf{landscape transferability} ensures those malicious inputs have high chance to fool the target model as well, especially when the shadow and victim models are transferred from the same PTLM. Black-box evasion attacks can be classified into two categories based on the granularity of adversarial perturbations, as described below. 


\vspace{3pt}
\noindent\textbf{Word-level attacks.}
The adversarial $\mathcal{U}$ can replace candidate characters or words with carefully-crafted ones to generate adversarial examples. There are two strategies to identify the adversarial worlds and characters. The first one is \textit{heuristic generation}, which designs the perturbation through some pre-defined rules. For example, \cite{jin2020bert} proposed TextFooler, a simple adversarial generation method to identify the critical words of text sentences that can affect the prediction of PTLMs. It first obtains the importance rank of each word by predicting the scores of the sentences 
without the word,  and then uses counter-fitting word embedding to semantically replace them until the prediction is altered. \cite{zang2019word} utilized the swarm optimization-based algorithm to search for sememe-based substitution words to mitigate the semantic warp to the original sentences. \cite{malik2021adv} presented Adv-OLM that adapts the idea of Occlusion and Language Models (OLM) to select candidate words for replacement.  \cite{maheshwary2021generating} leveraged the population-based optimization algorithm to improve the semantic indistinguishability of the generated adversarial examples. Instead of generating similar statements by synonym substitution, \cite{yin2020robustness} comprehensively summarized syntactically incorrect word generation methods that also threaten the robustness of PTLMs. In terms of the attack transferability, \cite{emmery2021adversarial} extended TextFooler with a  transformer-based replacement method that achieves high transferability.  \cite{yuan2021transferability} presented a systematic study to investigate the transferability of adversarial examples for text classification models and proposed a population-based genetic algorithm to find an optimal ensemble and highly-transferable adversarial words.

The second strategy is \textit{automatic generation}, which leverages an additional model to automatically generate substitution words to achieve better semantic indistinguishability. \cite{li2020bert} proposed BERT-Attack, which adopts pre-trained masked language models exemplified by BERT to generate adversarial examples. It computes the scores of sentences after replacing words with [MASK] to find the important words and then generates substitution words with BERT. Similarly, \cite{garg2020bae} presented BAE that utilizes contextual perturbations from a BERT masked language model to generate adversarial examples. Compared with BERT-Attack, BAE replaces and inserts tokens in original texts by masking a portion of the texts and leverages BERT-MLM to generate alternatives for the masked tokens. CLARE \cite{li2020contextualized} produces adversarial examples through a mask-then-infill procedure. It builds a pre-trained masked language model and modifies the inputs in a context-aware manner. \cite{shi2019robustness} leveraged a BERT masked language model to generate adversarial examples via modification with shared words.  In addition to using the language mask model for generating replacement words, MORPHEUS \cite{tan2020s} automatically perturbs the inflectional morphology of words to craft plausible and semantically similar adversarial examples.

\vspace{3pt}
\noindent\textbf{Sentence-level attacks.}
Instead of replacing certain words, these attacks craft adversarial sentences with better exploitation of sentence structures, contexts, etc. 
\cite{lin2021using} used irrelevant sentences to generate adversarial examples and attack machine reading comprehension (MRC) models. \cite{wang2020t3} proposed T3, a target-controllable adversarial attack framework. T3 uses a tree-based autoencoder to embed the discrete text data into a continuous representation space and then optimizes the adversarial perturbation. The tree-based decoder can generate adversarial examples directly from original sentences and regularize the syntactic correctness of the generated text. Such attack is also applicable for word-level adversarial example generation. \cite{gan2019improving} explored the robustness of question-answer (QA) models to question paraphrasing by creating two test sets. These two sets consist of paraphrased SQuAD questions that are generated using a neural paraphrasing model and re-writing the original question. \cite{zhang2019paws} introduced a paraphrase dataset, PAWS, which includes pairs of well-formed paraphrase and non-paraphrase with high lexical overlap to fully identify the robustness vulnerabilities of PTLMs.

\section{Privacy Threats}\label{sec:privacy}
Next we consider the privacy attacks against PTLM systems. The adversary does not affect the model or inference execution, but just passively sneak the critical information regarding the model or data. Such privacy threats could cause significant financial damage considering the DNN model or training data are becoming more valuable assets. 

\subsection{Data Privacy Attacks}
We first characterize attacks that compromise the privacy of training or inference data. Such attacks are launched by malicious $\mathcal{DSP}$ in the \textbf{fine-tuning} phase, or malicious $\mathcal{U}$ in the \textbf{inferring} phase. Recent studies \cite{feldman2020neural} have demonstrated that machine learning models can memorize data, which allows adversarial service providers or users to steal information from the model behaviors. Particularly for the PTLM scenario, the key information of training or inference samples may be extracted from the embedding codes and PTLMs \cite{song2020information,carlini2021extracting}. The \textbf{landscape transferability} makes the victim model more transparent to the adversary, and facilitates the attacks in the black-box setting. According to the type of extracted information, we classify data privacy attacks into the following three categories.

\vspace{3pt}
\noindent\textbf{Embedding inversion attacks.}
PTLMs are often used as embedding models to extract embedding representations
of words for downstream tasks. A malicious service provider can invert the original sentence of an inference input based on the corresponding embedding code. \cite{song2020information} designed such embedding inversion attacks in both white-box and black-box modes. In the white-box scenario, the embedding inversion attack is modeled as an optimization problem, and they proposed a continuous relaxation of sequential word inputs to achieve efficient optimization based on gradients. In the black-box scenario,
the adversary trains an inversion model that takes a text embedding code as input and outputs the set of words as the extracted results.

\vspace{3pt}
\noindent\textbf{Attribute inference attacks.}
One typical example of attribute inference attacks is membership inference attacks (MIAs), which aim to infer whether a data sample is in the training set from the target machine learning model. Existing MIAs \cite{shokri2017membership} have shown that given black-box access to the victim model, the predicted confidence of the test sample can reveal its membership attribute. Such attacks usually train a shadow model to distinguish between members and non-members of the training data. \cite{song2020information} developed a simple and efficient MIA based on the observation that the words or sentences in the context for training have more similar scores to each other than the ones in other contexts.

Another type of attribute inference attacks is keyword inference attacks, where the adversary is curious about whether certain keywords exist in an unknown inference sentence. The keywords can be highly sensitive and contain indicators for the adversary to extract private information, e.g., location, residence, medical records. Several works have designed such attacks against word embedding models. For example, \cite{pan2020privacy} trained attack models to predict the sensitive information of PTLMs from the embedding vector representations of word sequences, which can be applied in both the white-box and black-box modes. In the white-box scenario, the adversary is assumed to have a shadow corpus sampled from the same distribution as the unknown plain text, and trains a binary classifier with the dataset to distinguish whether this keyword is in plain text or not. In the black-box scenario, the adversary has little prior knowledge of the plain text and thus obtains an external corpus from other public corpora. To mitigate the domain misalignment phenomenon, an additional module is introduced
to help transfer the adversarial knowledge.
\cite{song2020information} also designed a similar attack to infer the sensitive attributes of a word sequence using its embedding vector representation. In their threat model, the adversary only has limited labeled corpus so as to closely match real scenarios where labeled sensitive corpus is hard to be collected.

\vspace{3pt}
\noindent\textbf{Corpus inference attacks.}
In addition to the recovery of inference data, recent research also shows the feasibility of extracting the training corpus from PTLMs or downstream models. \cite{carlini2021extracting} used the perplexity of the generated sequences to choose top 100 sequences as the extracted training data of publicly available PTLMs. \cite{zanella2020analyzing} analyzed the information leakage in fine-tuned downstream models and proposed a differential score metric to capture the difference between probabilities assigned to word sequences by PTLMs and their fine-tuned version. The word sequences with higher differential scores belong to the private training corpus.
Instead of reconstructing the corpus for training PTLMs, \cite{panchendrarajan2021dataset} aims to reconstruct the corpus for fine-tuning the downstream tasks. The objective of the adversary is to iteratively analyze the difference between PTLMs and their fine-tuned models to identify sentences belonging to the fine-tuning corpus.

\subsection{Model Privacy Attacks}\label{sec:ip}
The PTLM and downstream models are very important intellectual property, and deserve well protection. However, a malicious user could perform model extraction attacks (MEAs) to reconstruct the proprietary model by just remotely querying the system in the \textbf{inferring} phase. This is enabled by the high \textbf{landscape transferability}. According to the extraction goals, we can classify existing MEAs into two categories \cite{jagielski2020high}.


\vspace{3pt}
\noindent\textbf{Accuracy extraction attacks.}
This type of MEAs intend to extract a model with similar accuracy on the text data as the victim PTLM. For instance, \cite{he2021model} proposed to extract BERT-based models using a task-specific query generator to construct word sequences as queries. With the extracted model, they also utilized the transferability to conduct attribute inference and adversarial attacks. \cite{zanella2021grey} designed an algebraic extraction attack against BERT-based models in the grey-box setting. The encoder is frozen during fine-tuning and the adversary is allowed to compute hidden embeddings of text inputs by querying a public encoder model. The adversary first distills an initial model based on the public encoder and then uses the resulting encoder to extract and replace the last layer.

\vspace{3pt}
\noindent\textbf{Fidelity extraction attacks.}
This type of MEAs aims to steal a PTLM with similar behaviors as the victim one. To achieve this goal, \cite{xu2021beyond} accomplished fidelity extraction through an imitation attack. The adversary learns an imitation model based on the queries and retrieved labels from the victim model to enhance the behavior similarity. The combination of the domain adaptation theory and multi-victim ensemble methodology is also adopted to boost the attack performance. \cite{keskar2020thieves} extracted multilingual models through querying gibberish data to a monolingual model. The task-specific knowledge can be extracted with high accuracy on several languages. The extraction performance can be improved if the adversary has access to the real data.

\section{Discussion and Conclusion}\label{sec:conclusion}
In this paper, we present a comprehensive survey about the security threats and attack surface in the pre-trained language model scenario. We characterize the threats from three novel angles: system pipeline, attack goal and model transferability. We summarize existing state-of-the-art attacks against PTLMs, including backdoor and evasion attacks for integrity, model and data privacy attacks. We hope our survey can raise people's attention to the PTLM security in both academia and industry, and help them better understand and identify the underlying mechanisms of those potential threats. 

Inspired by the above survey, we present several open problems and potential research directions for both attacks and defenses in PTLM systems:

\vspace{3pt}
\noindent{\textbf{Robustness enhancement}}. 
There exists an arms race between the integrity attacks and defenses. From the adversary's point of view, a variety of backdoor detection or removal methods \cite{qi2021onion,fan2021defending} have been proposed to guard standalone models, and they are possibly effective for PTLMs as well. It is interesting to design more sophisticated attacks to evade such defenses. From the defender's point of view, how to combine the characteristics of PTLM systems with conventional robustness solutions (e.g., adversarial training \cite{liu2020adversarial}) to improve the model's robustness and generalization is worth exploration.


\vspace{3pt}
\noindent{\textbf{Trade-off between utility and security}}. One common technique of preventing information leakage is to obfuscate the model parameters or inference behaviors. For instance, \cite{pan2020privacy} added Gaussian noise to defeat membership inference attacks. However, such defense strategy would affect the utility of the PTLMs. 
How to achieve better trade-off between the model performance and defense effectiveness remains an open problem. 

\vspace{3pt}
\noindent{\textbf{Transferability improvement}}. As discussed before, transferability is an inherent feature of PTLMs. It maximizes the generalization ability of PTLMs, but also facilitates some attacks. It is interesting to study how to reduce the attack transferability, while maintaining the model's generalization. For instance, for backdoor threats, it is valuable to devise a fine-tuning method that only transfers the knowledge of the PTLM for normal data while forgetting the knowledge of malicious data with triggers. For privacy, how to make the target model less replicable for further attacks is worth investigation. 

\clearpage
\bibliographystyle{named}
\bibliography{ref}

\end{document}